# Optimization of efficiency and energy density of passive micro fuel cells and galvanic hydrogen generators

Robert Hahn, Stefan Wagner, Steffen Krumbholz, Herbert Reichl,
Fraunhofer IZM, Gustav- Meyer-Allee 25,
D-13355 Berlin, Germany

*Abstract*- A PEM micro fuel cell system is described which is based on self-breathing PEM micro fuel cells in the power range between 1 mW and 1W. Hydrogen is supplied with on-demand hydrogen production with help of a galvanic cell, that produces hydrogen when Zn reacts with water. The system can be used as a battery replacement for low power applications and has the potential to improve the run time of autonomous systems. The efficiency has been investigated as function of fuel cell construction and tested for several load profiles.

## I. INTRODUCTION

During the last few years, the development effort related to small, portable fuel cells has increased significantly. The main motivation underlying the development of micro fuel cells is the possibility to achieve higher energy densities compared to batteries.

This development benefits greatly from the existing knowledge and attempts to improve larger fuel cells for automotive, residential, and stationary applications. For the commercialisation of both big and small fuel cell systems, however, improvements still are required in several areas. For DMFCs, for example, it has been recognised that the success of this fuel cell technology depends largely on developing better membranes with lower methanol cross-over and improving the electro-catalysts which can overcome the slow anode kinetics.

When developing smaller fuel cells, it is impossible to simply use scaled-down systems architectures and components applied in their larger counterparts.

A complete portable fuel cell system consists of three major parts:

The fuel cell stack which is the core of the system. Its size is related to the power output.

The fuel tank. Its size is related to the amount of stored energy and, hence, to the runtime of the device.

The balance of plant (BOP) which includes all the peripheral components that support the power generation process. In most cases, this is the hydrogen-generating system for PEM fuel cells.

Since compressed gas or liquid hydrogen cannot be used for portable or small fuel cells, the research focuses on three kinds of fuel cell. The first are the direct liquid fuel cells using methanol (DMFC), ethanol (DEFC) or formic acid (DFAFC). Then, the PEM fuel cells with hydrogen are considered, where the hydrogen is generated from reformed methanol, reversible storage alloys or chemical hydrides and water-reactive alloys. The last type of interest, but still in the state of basic research is the biofuel cell. In this case, organic materials like alcohols, organic acids or glucose are used as a fuel and biocatalysts convert chemical into electrical energy. So far, the existing prototypes have shown a very low power density and short lifetime. Therefore, they will not be examined here.

Many attempts have been made to date to reduce the balance of plant of portable fuel cells in order to increase reliability and reduce costs. The related studies revealed that for portable fuel cells sophisticated peripheral components have to be developed to allow for a higher power density and operation under varying loads and ambient conditions. The key challenge in this field is how to achieve the desired power performance, while simplifying the design of the BOP in order to miniaturise the whole system. With miniaturisation, application-specific components like valves and pumps based on micro systems technology have to be developed.

*A. The fuel cell core – micro fabrication technologies*

Typically, large fuel cells are mechanically compressed sandwiches of a graphite composite or metal electrodes and membrane assemblies. Each component of the fuel cell has to be re-designed based on well-established technology platforms for miniature components in order to achieve a cost-effective miniaturisation. Therefore, most researchers use available manufacturing techniques like:

- Silicon and MEMS technologies

   

- Foil processing of polymer and metal foils, polymer substrates
- Printed-circuit board technology
- Planar ceramic technology like low-temperature co-fired ceramics (LTCC).

The advantage of printed-circuit board technology as a basis of flow field and current collector fabrication above all is the low-cost mature technology. Furthermore, light-weight and stiff composite materials are used and design flexibility is ensured, as complex conductor/insulator patterns are applied either as a mono- or multi-layer design. However, the standard material like the copper/glass epoxy composite of printed-circuit boards cannot be used for long-term stable fuel cells without modification. Indeed, copper would degrade in the corrosive fuel cell environment. This, in turn, will increase not only the contact resistance between electrode and current collector, but also introduce metal ions that may enter the membrane and significantly decrease the membrane conductivity.

That is why appropriate material modifications or surface coatings have to be established in order to use printed-circuit board technology in portable fuel cell applications.

The majority of research activities related to micro-scale fuel cells is also aimed at micro-power applications. There are many new miniaturised applications which can only be implemented, if a higher-energy-density power source is available compared to button cells and other small batteries. Miniaturisation of the conventional fuel cell stack technology is not possible down to these dimensions.

*B. Hydrogen supply*

Pressure tanks, liquid hydrogen or gas/ methanol reformers are not practical solutions for micro systems. Therefore a galvanic cell was used to generate hydrogen on demand. Thus, the combination of micro hydrogen generator and planar micro fuel cell results in a highly compact and reliable system.

II. FUEL CELL DEVELOPMENT

*A. Technology*

At Fraunhofer IZM the use of a variety of thin-film and printed circuit board substrates as a basis of flow field and current collector fabrication was investigated. The assembly technology for micro fuel cells based on electronics manufacturing and reel-to-reel processing for mass production was developed together with industrial partners [1].

For the fabrication of hydrogen fed planar passive PEM (polymer electrolyte membrane) fuel cells with commercial MEAs (membrane electrode assembly) in the size range between 1 mm and several centimetres, three technological approaches were investigated:

- Thin-film metallization on patterned polyimide and metal foils, no gas diffusion layers (GDL),
- Printed circuit boards with electroplated and etched micro channels, no GDL,
- Printed circuit boards with electroplated and etched micro channels in combination with GDLs.

Two approaches were studied for the openings of the cathode side, which are required for air and water exchange with the environment: Lamination of foil substrates with current collector metal lines and isolating substrates with slots rectangular to the metal lines. Alternatively substrates with drilled holes were used and metallizations surrounding the holes.

The basic planar fuel cell design is shown in Fig. 1.

Fig.2 shows anode flow fields of the thin film type cells while flow fields based on printed circuit board technology are demonstrated in fig. 3.

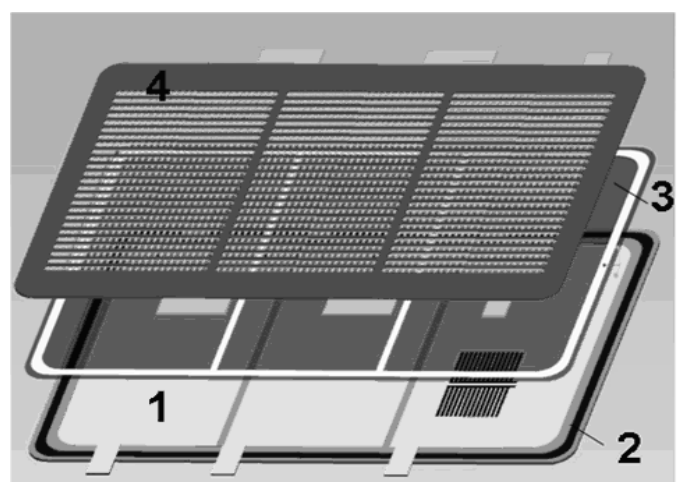

Fig. 1. Schematic of a planar micro fuel cell. 1: anode flow field, 2: sealing of the hydrogen domain to the membrane 3 (Membrane Electrode assembly, MEA), 4: Cathode current collector.

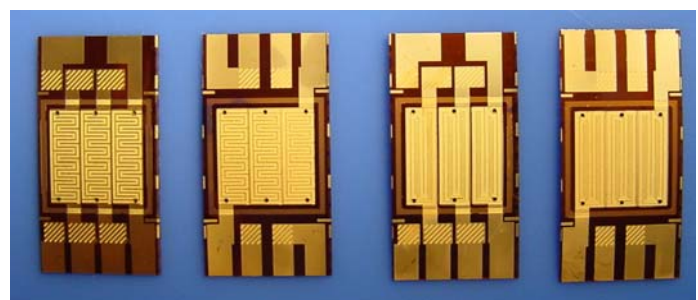

Fig.2. Anode side of thin film-foil type micro fuel cells (DF).

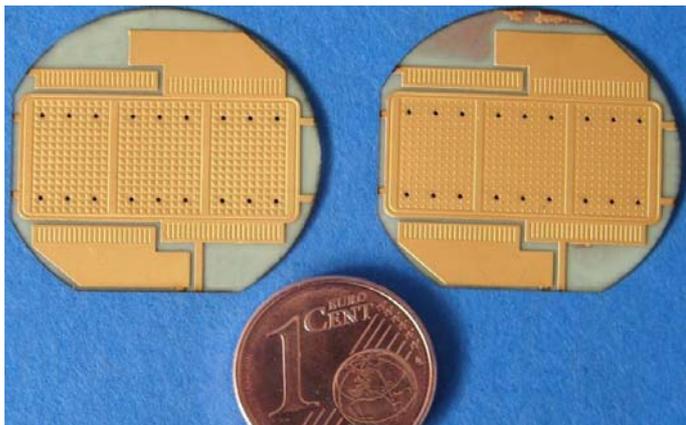

Fig. 3. Anode flow fields fabricated in printed circuit board technology (PCB).

An overview of dimensions and technology is summarized in Table I.

TABLE I
INVESTIGATED FUEL CELLS

| Type | DF | PCB | PG |
|---|---|---|---|
| active area, A [cm$^2$] | 0.5 | 2 | 10 |
| technology | thin film/flex | PCB | PCB |
| Current collector structures, pitch p [µm] | 40 …800 | 600 | 500 …2000 |
| Use of GDL | - | - | x |

The technologies were evaluated in terms of:
- smallest possible fuel cell dimensions,
- minimum channel dimensions and the option to avoid gas diffusion layers,
- achievable ratio of air openings at the cathode side,
- stiffness of the substrates for uniform contact resistance,
- lateral electrical conductivity in current collectors for low ohmic losses,
- hydrogen leakage which would reduce efficiency in the low power mode,
- corrosion and long term-stability,
- capability for volume manufacturing and costs.

The influence of structure dimensions on fuel cell performance and trade-offs between manufacturing cost and fuel cell internal losses have been studied in detail. Significant cost savings can be achieved, when established manufacturing technologies can be used.

Assembly of the fuel cells was performed with the help of screen printed gaskets and adhesive dispensing for electrical contacts. Processing temperatures must be below ca. 100°C to avoid degradation of the ionic membrane. Another issue was the handling of very thin membranes down to a thickness of 10 µm.

Several design variations were tested for each technology. A maximum power density of 150 … 200 mW/cm$^2$ was obtained.

The electronic substrate technologies allow the simultaneous fabrication of planar stacks with several interconnected cells.

*B Design optimization and characterization*

As only commercial membrane electrode assemblies (MEA) are used, the efficiency can be controlled only marginally. The Faraday efficiency is dependent on the amount of hydrogen diffusion through the membrane. Sealing of the anode space must be carried out such, that leakage through the sealing is much lower compared to the membrane diffusion.

The voltage efficiency can be reduced by improper design of the current collectors. In contrast to the conventional fuel cell stack, the current is conducted laterally through the current collectors to the fuel cell terminals. If diffusion layers are not used, the lateral conduction in the catalyst layer can cause significant losses as well.

In general, the resistivity of electron current is the sum of the in plane ($R_i$) and through plane resistance of the catalyst / GDL layer ($R_t$), the contact resistance between catalyst /GDL ($R_c$) and current collector and the resistance of the metallic current collector ($R_m$) on both, anode and cathode side.

$$R_s = R_i + R_t + R_c + R_m \qquad (1)$$

To demonstrate the effect of ohmic losses on planar fuel cells, the influence of the total electrical resistivity on the V/I-curve, the efficiency the power output and the relative power loss is shown in Fig. 4 …7. All electrical conduction losses are summarized in the resistance $R_s$ which is normalized to one square centimeter of active fuel cell area.

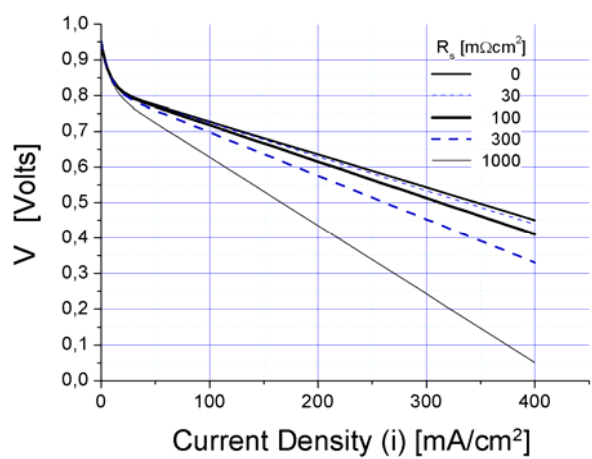

Fig. 4. Influence of Ohmic losses on V/I curve of a passive planar fuel

cell.

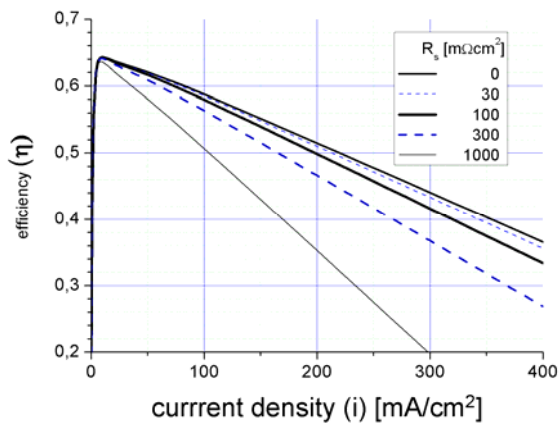

Fig. 5. Influence of Ohmic losses on the efficiency of a passive planar fuel cell.

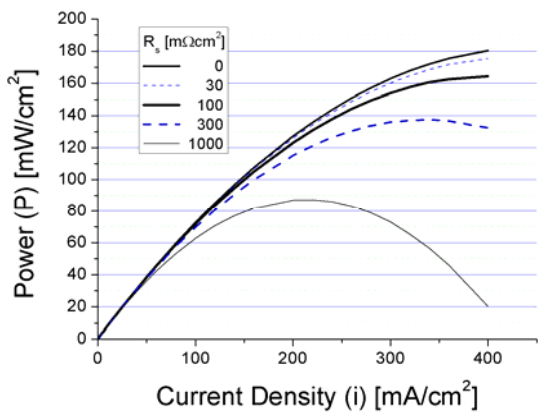

Fig. 6. Influence of Ohmic losses on the power output of a passive planar fuel cell.

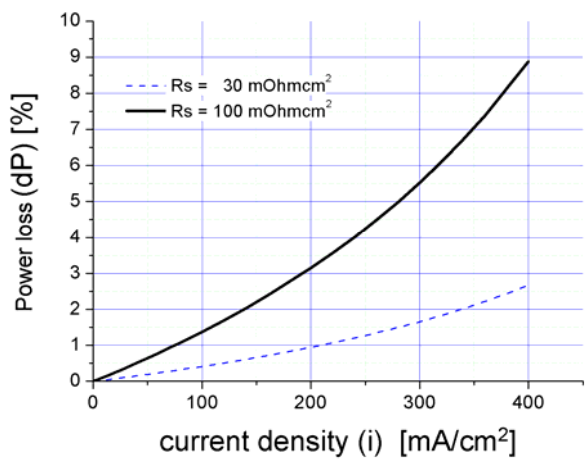

Fig. 7. Resulting relative power reduction of a passive planar fuel cell due to ohmic losses.

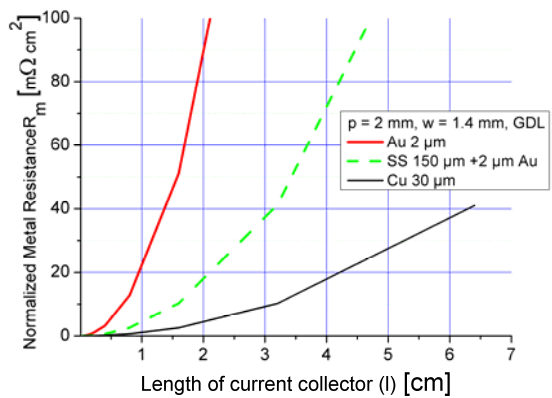

Fig. 8. Normalized metal resistance as function of length for cathode current collector with 0.6 mm wide ribs at 1.4 mm distance.

Power and efficiency losses can be kept in the one percent region, if $R_s$ stays below ca. 100 mOhm. If the working point is at low currents in the high efficiency region, even higher electrical resistances are acceptable. The electrical resistances of the current collectors used are shown in Fig.8 for an opening ratio of 70 %. Electroplated gold, which is used in the thin film-foil design (type DF) should not be longer than ca. 1 cm. The longest current collectors can be fabricated with copper (30 µm Cu, in printed circuit board technology, type PCB) while 150 µm gold plated stainless steel mesh, which is used as current collector for type PG, should be limited to ca. 2 cm in length.

A straight forward one dimensional resistance model was used to calculate the proportion of the four resistance contributions of equation (1) according to the planar fuel cell design [2]. The model parameters are summarized in Table II.

TABLE II
PARAMETERS FOR RESISTANCE CALCULATION

| | |
|---|---|
| $\rho_I$, in plane resistivity, GDL | 50 mΩcm |
| $\rho_T$, through plane resistivity, GDL | 400 mΩcm |
| $\rho_I$, in plane resistivity, catalyst layer, MEA | 360 mΩcm |
| $\rho_T$, through plane resistivity, catalyst layer, MEA | 360 mΩcm |
| $t_{GDL}$, thickness GDL | 325 µm |
| $t_{cat}$, thickness catalyst layer | 10 µm |
| $\rho_c$, contact resistance between current collector and GDL /MEA | 4 mΩcm² |

The results for the cathode side of the investigated fuel cell types according to table I are shown in Fig. 9. The channel pitch was varied at a constant opening ratio of 60 %. Here fuel cells with GDL (type PG) has higher ohmic losses compared to the other prototypes. But thinner GDLs with higher conductivity can be used alternatively to reduce in-plane and through plane resistances. To achieve low in-plane resistances, the pitch should be held below 400 µm and 2 mm for types without and with GDL respectively.

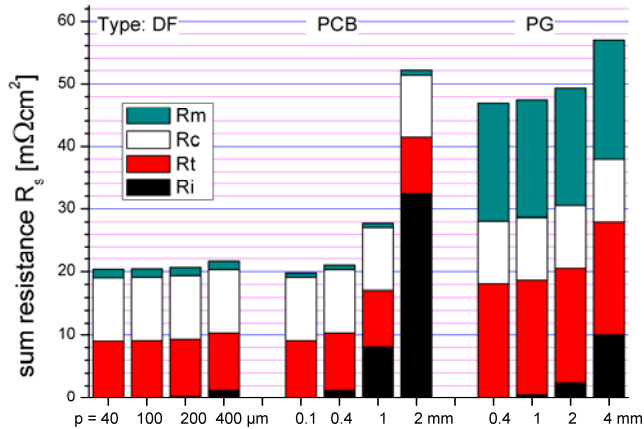

Fig. 9. Overview of electrical resistances of cathode side electron transport of the investigated fuel cell types according to Table I. Variation of the channel pitch p, open ratio of 60 %.

Another significant resistance contribution comes from the stainless steel current collectors. They are covered with 2 µm gold to reduce the contact resistance. Long term experiments have proven, that only stainless steel or pure Au current collectors (type DF) are sufficiently stable. Ni-Au-plated copper lines (type PCB) therefore are not the right alternative.

To investigate the influence of the GDL, a comparison was made of fuel cells of type PCB with and without GDL and type DF. As can be seen in Fig. 10a there are only minor differences in the polarization curves. The same MEA has been used for all types. If a GDL is used, the voltage is a bit lower at low current densities. At high current densities the voltage of type DF and PCB drops faster. The fuel cell with GDL shows the highest power density. This results probably from better water management [5].

The polarization curves and power density of two planar fuel cells of type PG are shown in fig. 10b. These fuel cells with GDL show a slightly more pronounced voltage reduction at low current density but more stable curves at high power density.

In summary, the ohmic losses are sufficiently low for all fuel cell types investigated.

Electrical conduction is only one part of the design optimization. If the current collector ribs are too wide, the oxygen concentration under the rib can reduce significantly. Numerical simulation has shown, that the oxygen mol fraction within the catalyst layer at 0.6 V is reduced to below 1 % at a lateral distance from the channel of ca. 25 µm if no GDL is used [3] and at a distance of ca. 400 µm at 800 mA/cm$^2$ with a 200 µm thick GDL [4].

The removal of reaction water is also influenced by the rib width. Therefore large current collector ribs can reduce the fuel cell performance. The channel width w and pitch p as well as the opening ratio w/p have to be optimized.

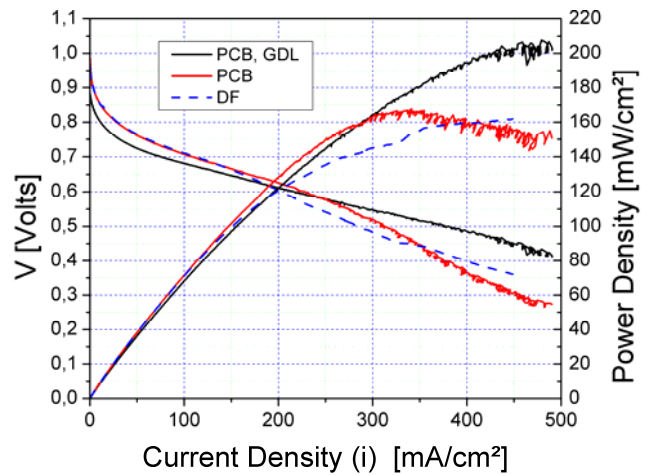

Fig. 10a. Measured polarization curve and power density of planar fuel cells (Types DF and PCB) according to Table I

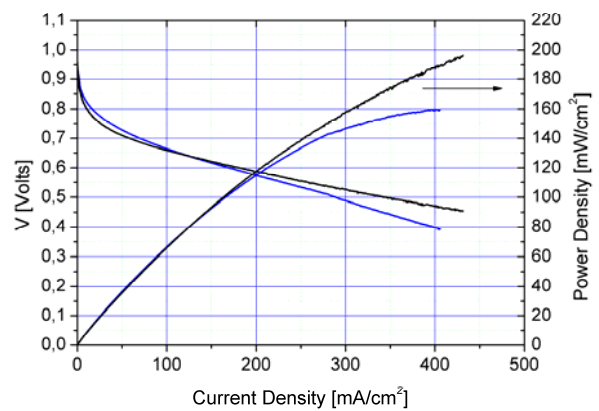

Fig. 10b. Measured polarization curve and Power density of planar fuel cells type PG according to Table I

The electronic substrate technologies – either foil-type or printed circuit board - allow the simultaneous fabrication of planar stacks with several interconnected cells. Fuel cell assembly can be improved substantially if the same electrolyte membrane is used for all cells. The electrical isolation space between adjacent cells must be sufficiently high to avoid leakage currents between adjacent cells which would reduce the Faraday efficiency.

*C. Fuel cell efficiency*

While most authors are focussing on increasing the peak power density, which is important for portable electronics applications like mobile phones, where the power requirements are rather high, for other applications like autonomous sensors and other long running devices fuel cell efficiency is of prime interests. This means high voltage output at medium currents and low hydrogen losses.

At higher current density, the fuel cell efficiency is dominated by the over-potentials while at low current densities hydrogen leakage dominates. The efficiency can be approximated as the product of Faraday efficiency and voltage efficiency according to (2).

$$\eta = \eta V * \eta F = V/V_{ref} * I/(I+I_{leak}) \quad (2)$$

The influence of the leakage current is demonstrated in Fig. 11. The current voltage characteristic of Fig. 4, Rs =0 was used in equation (2) with $V_{ref}$ = 1.23 Volts.

The fuel cell efficiency was measured with help of a test set up which allows precise measurement of sub µl hydrogen consumption of the fuel cell. This is shown for three planar fuel cells of the PCB-type in fig. 12. Measurement curves as function of fuel cell currents were fitted to estimate the hydrogen leakage of the system. The Faraday efficiency corresponds to 0.4 … 0.7 mA/ cm$^2$ of membrane area. This is in good agreement with the data sheet values of the used MEA.

Serial interconnected cells with only 200 µm electrode space showed a reduced efficiency (sample PB-08, Fig.12). Maximum efficiency of ca. 56 % was achieved at a power density of 20 mW/cm$^2$. Experiments with pulsed load (70 …220 mW, 7 msec, 100 msek period) showed ca. two percent lower efficiency compared to a constant load of 5 mW. At higher duty cycles the efficiency drops because the hydrogen leakage becomes the dominant loss mechanism.

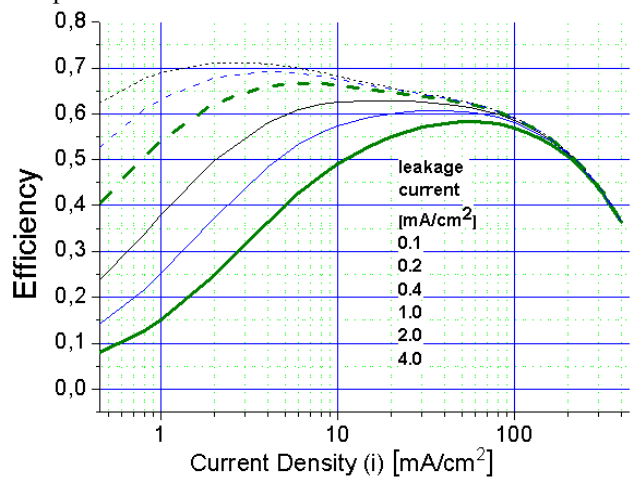

Fig. 11. Fuel cell efficiency as function of leakage current density according to (2)

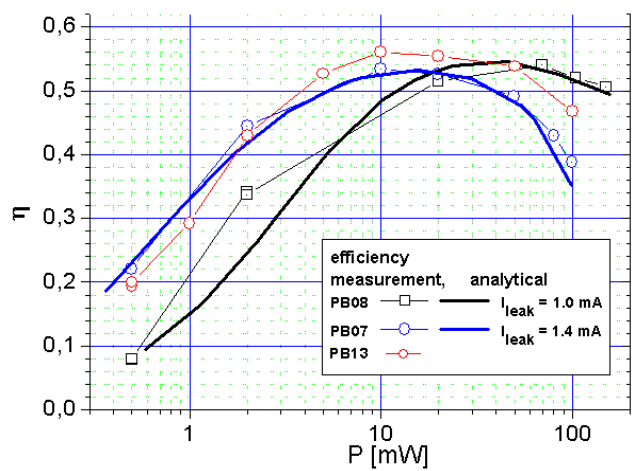

Fig. 12. Measured fuel cell efficiency for type PCB as function of power

III. HYDDROGEN GENERATOR AND SYSTEM INTEGRATION

Since hydrogen pressure tanks are not practical for small electronic devices or micro systems a galvanic cell was developed for the hydrogen production on demand. Hydrogen is produced by the reaction of zinc and water according to equation (3).

$$Zn + H_2O \rightarrow ZnO + H_2 \quad (3)$$

Since the hydrogen flow is proportional to the cell current according to Faraday's law, the rate control is straightforward. If the fuel cell and the galvanic cell are electrically connected in series, the galvanic cell produces precisely the amount of hydrogen the fuel cell consumes. Only minor leakages have to be compensated.

Thus, a very low component overhead compared to other known hydrogen-generating systems can be achieved, since pressure or flow controllers and valves are not required.

Currently, hydrogen-evolving cells are commercially available only in the form of a button cell, as shown in Fig. 13 and Fig.14.

A zinc cell is composed of the Zn electrode and KOH electrolyte very similarly to a classic primary zinc-air cell and has an equivalent configuration.

The cell consists of a gas-generating electrode, a counter electrode with the active mass and filled with an aqueous electrolyte. The gas electrode and the counter electrode are electrically connected to the metal cup and metal cover, respectively, to provide for the electrical contact to the outside.

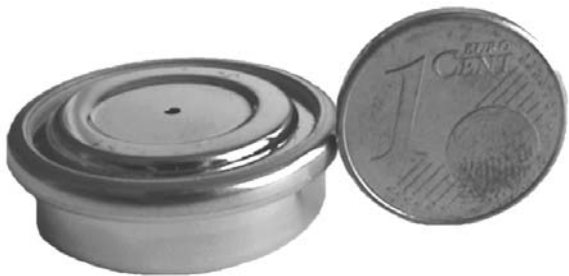

Fig. 13. Button cell hydrogen generator, size 3.5 cm³.

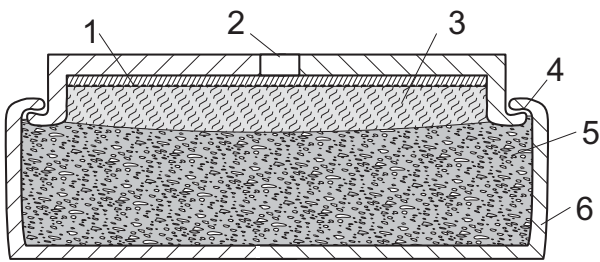

Fig. 14. Cross-sectional view of a hydrogen-generating button cell: 1- gas diffusion electrode with catalyst, separator; 2- opening; 3- water/electrolyte; 4- plastic seal; 5- Zn electrode; 6- metal cover.

The main difference from primary batteries is that the system has an opening for hydrogen release. In the past, these systems were developed for lubricant dispensers, drug delivery or hydrogen reference electrodes. The user will replace the hydrogen cell like a primary battery. The costs are anticipated to be similar to those of button-sized cells or alkaline batteries.

As stated above, additional components, such as pressure regulators and valves, are not required. This reduces costs and complexity. Moreover, system efficiency and miniaturisation can be improved dramatically.

Since the hydrogen flow is proportional to the cell current according to Faraday's law, the most straightforward electrical circuit is a serial interconnection. Fig. 15 displays an electrical circuit of a hydrogen generation cell and a fuel cell system. The Schottky diode acts as a current by-pass during the start phase when a load current would result in a reverse fuel cell polarity. Once enough hydrogen is produced to establish a fuel cell potential, the diode is blocked. It should be noted that not only the fuel cell, but also the galvanic hydrogen cell adds to the system voltage. Depending on the current, the hydrogen-evolving cell yields a voltage between 0 and 0.4 volts. The by-pass resistor $R_L$, in parallel to the gas-evolving cell, is designed to compensate hydrogen losses or any deviation from a current efficiency equal to one. Instead of the resistor $R_L$, a controllable electronic load can be used to improve hydrogen generation at load steps.

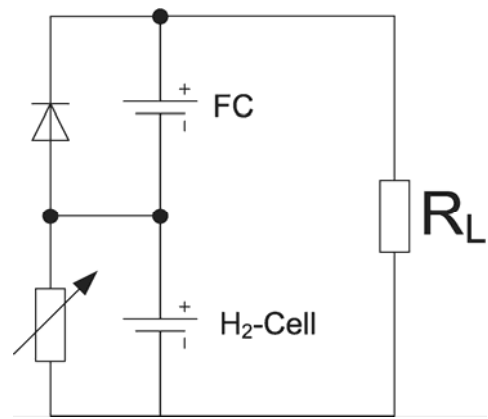

Fig. 15. Electrical circuit of a fuel cell with hydrogen generation cell.

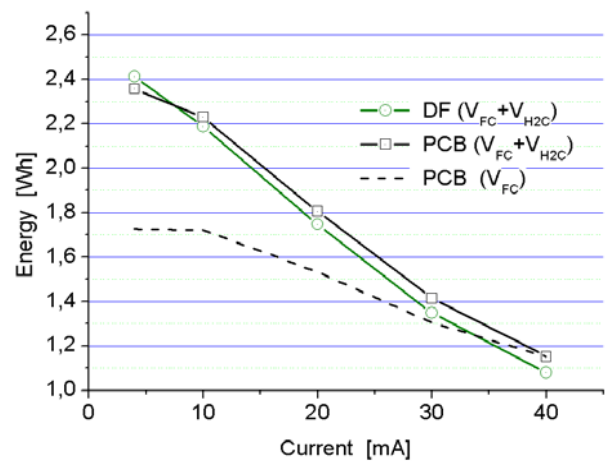

Fig. 16. Obtainable energy of 4 cm³ fuel cell system with on demand hydrogen generator.

A major disadvantage of small fuel cells with galvanic hydrogen generation cells is the low hydrogen rate and the resultant low power density of the system. In comparison to alkaline batteries, the energy density in terms of the generated hydrogen volume is reduced significantly at higher currents. The prototype system of 4 cm$^3$ in size was designed for a power output between 5 and 20 mW. The energy reduction with increasing current is shown in Fig. 16. At 5 mA the small fuel cell (DF) yields a somewhat higher efficiency. The dashed line shows the energy obtainable if only the fuel cell voltage is considered.

Several adapters which integrate fuel cells of type DF and PCB and the hydrogen generation cells are shown in Fig. 17.

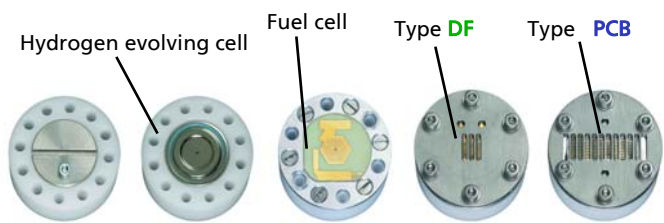

Fig. 17. Fuel cell test adapter with hydrogen generation cell.

TABLE III
EFFICIENCY OF MICRO FUEL CELLS AS FUNCTION OF DUTY CYCLE, 70 mW PULSES, 7msec

| Duty cycle | Interval | mean power | η DF fuel cell A = 0.5 cm$^2$ | η PCB fuel cell A = 2.0 cm$^2$ |
|---|---|---|---|---|
| 1 /10 | 100 ms | 4.77 | 0.65 | 0,4 |
| 1 / 100 | 1 sec. | 0.5 | 0.55 | 0.09…0.2 |
| 1 / 1000 | 10 sec. | 0.07 | 0.15 | - |

The size of the hydrogen cell and the fuel cell has to be properly adjusted. If the fuel cell membrane area is too large, the amount of hydrogen that diffuses through the membrane becomes significant. This reduces the overall efficiency during idle mode or low current operation of the electronic device. If the fuel cell area is too small, than the fuel cell voltage drops at high current density and the fuel cell efficiency decreases. Therefore, the fuel cell size has to be optimized, depending on the load profile and the size of the hydrogen evolution cell. This is demonstrated in table III.

At higher duty cycles a hybrid system should be used where the fuel cell charges an electrical buffer like a super capacitor or a lithium secondary battery at a constant current. For a mean power of 0.5 and 0.07 mW an active fuel cell area of only 2.5 and 0.35mm$^2$ is required respectively.

Due to electronic manufacturing technology such small dimensions and design changes of the fuel cell according to the systems specifications can be made easily.

CONCLUSIONS

A micro PEM fuel cell in combination with $Zn/H_2O$ hydrogen production was described.

A design optimization of planar passive fuel cells was carried out based on thin film-foil and printed circuit board technologies.

At first time an efficiency evaluation of micro fuel cell systems at low power and pulsed discharge as been carried out. This system may be suitable as power supply for typical radio sensor nodes.

ACKNOWLEDGMENT

This project was in part funded by the German Ministry of research and education. Hydrogen evolving cells were supplied by Varta Microbattery GmbH.

REFERENCES

[1] S.Wagner, R. Hahn, R. God, H.-P. Monser, *Low Cost Manufacturing of foil-type micro fuel cells*, mst news No. 4/05 August 2005, pp.34-36
[2] Robert Hahn, Stefan Wagner, Steffen Krumbholz, Herbert Reichl, *Development of micro fuel cells with organic substrates and electronics manufacturing technologies*, to be published at ECTC 2008, May 27-30, 2008
[3] Stefan Wagner, Thesis, TU-Berlin, 2008
[4] Tianhong Zhou, Hongtan Liu, *Effects of the electrical resistances of the GDL in a PEM fuel cell*, Journal of Power Sources 161 (2006) 444-453
[5] Tibor Fabian, Jonathan D. Posner et. Al*., The role of ambient conditions on the pwerformance of a planar, air-breathing hydrogen PEM fuel cell,* Journal of Power Sources 161 (2006) 168-182